\documentclass{aa}  
\usepackage{graphicx}
\usepackage{array}    
\usepackage{amssymb}    
\usepackage{url}
\usepackage{color}
\usepackage{multirow}
\usepackage{makecell}

\usepackage{adjustbox}

\usepackage{txfonts}

\def\tef{$T_{\rm eff}$}
\def\vmic{$V_{\rm mic}$}
\def\vsin{$v \sin i$}
\def\lgg{$\log g$}

\def\msun{$M_{\odot}$}
\def\rsun{$R_{\odot}$}

\begin{document} 

   \title{50 Dra: Am-type twins with additional variability in a non-eclipsing system }


   \author{M. Skarka
          \inst{1}
          \and
          J. Lipt\'{a}k\inst{1,2}
          \and
          E. Niemczura\inst{3}
          \and
          Z. Mikul\'{a}\v{s}ek\inst{4}
          \and
          M. Cabezas\inst{1,2}
          \and
          M. V\'{i}tkov\'{a}\inst{1,4}
          \and
          R.~Karjalainen\inst{1}
          \and
          P.~Kab\'{a}th\inst{1}
          }
   \institute{Astronomical Institute of the Czech Academy of Sciences, Fri\v{c}ova 298, CZ-25165 Ond\v{r}ejov, Czech Republic\\
              \email{skarka@asu.cas.cz}
        \and
            Institute of Theoretical Physics, Faculty of Mathematics and Physics, Charles University, V Hole\v{s}ovi\v{c}k\'{a}ch 2, 180 00 Praha 8, Czech Republic
         \and
            Instytut Astronomiczny, Uniwersytet Wrocławski, Kopernika 11, PL-51-622 Wrocław, Poland
        \and
            Department of Theoretical Physics and Astrophysics, Masaryk University, Kotl\'{a}\v{r}sk\'{a} 2, CZ-61137 Brno, Czech Republic
             }

   \date{Received July xx, 2023; accepted July xx, 2023}

 
  \abstract
   {The interplay between radiative diffusion, rotation, convection, and magnetism in metallic-line chemically peculiar stars is still not fully understood. Recently, evidence has emerged that these effects can work together.}
   {Our goal is to study the bright binary system 50\,Dra, describe its orbit and components, and study additional variability.}
   {We conducted our analysis using TESS short-cadence data and new high-resolution spectroscopic observations. We disentangled the spectra using \textsc{Korel} and performed spectral synthesis with \textsc{Atlas9} and \textsc{Synthe} codes. The system was modelled using \textsc{Korel} and \textsc{Phoebe2.4}. We also employed SED fitting in \textsc{Ariadne} and isochrone fitting using \textsc{Param1.5} codes.}
   {Our findings indicate that the non-eclipsing system (with an inclination of 49.9(8)\,deg) 50\,Dra, displaying ellipsoidal brightness variations, consists of two nearly equal A-type stars with masses of $M_{1}=2.08(8)$ and $M_{2}=1.97(8)$\,\msun\, and temperatures of 9800(100) and 9200(200)\,K, respectively. Our analysis also suggests that the system, with an orbital period of $P_{\rm orb}=4.117719(2)$\,days, is tidally relaxed with a circular orbit and synchronous rotation of the components. Furthermore, we discovered that both stars are metallic-line Am chemically peculiar stars with an underabundance of Sc and an overabundance of iron-peak and rare-earth elements. We identified additional variations with slightly higher frequency than the rotational frequency of the components that we interpret as prograde g-mode pulsations.
   }
   {The system 50\,Dra exhibits numerous exciting phenomena that co-exist together and may have an impact on our understanding of chemical peculiarity and pulsations.
   }

    \maketitle
%

\section{Introduction}\label{Sect:Introduction}

About a third of spectral A-type stars show a deficiency of He, Ca, and/or Sc an overabundance of iron-group and rare-earths metals \citep{Abt1981, Gray2016}. These stars, which are mostly observed among stars with spectral type earlier than F2 within a typical temperature range between 7250 and 8250\,K \citep{Gray2016, Qin2019}, are called metallic-line chemically peculiar (CP) stars, shortly AmFm stars. The peculiar chemical composition is enabled by atomic diffusion, which occurs in stars with stable outer layers that transfer energy through radiation \citep{Michaud1970}. This condition is satisfied in slowly rotating stars ($<\simeq100$\,km\,s$^{-1}$) where rotational mixing is weak \citep{Abt1995, Qin2021, Trust2020}. It is not surprising that more than 70\,\% of AmFm stars are found in binary systems, especially those with orbital periods shorter than 20\,days, with a peak at around 5\,days, where tidal effects slow down the stars' rotation rate \citep{Abt1961, Abt1985, Carquillat2007}. It appears that systems with both components being of AmFm type are quite frequent. \citet{Catanzaro2024} studied six eclipsing binaries with AmFm stars and found that four of these systems exhibit the AmFm peculiarity in both primary and secondary components.

Since He is expected to quickly gravitationally settle down in AmFm stars \citep{Charbonneau1991}, they have not been expected to pulsate. However, there were examples of Am stars showing p-mode pulsations even before the availability of ultra-precise space data \citep{Kurtz1989}. Currently, it is clear that AmFm stars can pulsate as $\delta$\,Sct, $\gamma$\,Dor and hybrid pulsators \citep[e.g.][]{Balona2015,Smalley2017,Durfeldt-Pedros2024}. It was also discovered that if there is not sufficient He in the He\,II ionization zone, the pressure modes in $\delta$ Sct AmFm stars can be either excited by the turbulent pressure mechanism in the hydrogen ionization zone \citep{Antoci2014, Smalley2017} or by a bump in Rosseland mean opacity resulting from the discontinuous H-ionization edge in bound-free opacity \citep{Murphy2020}.

It is usually assumed that rotationally induced variability can only be observed in CP stars with strong (kG), globally-organized magnetic fields that can stabilize abundance spots \citep[Ap/Bp stars,][]{Preston1974}. However, a recent investigation of an Ap star 45\,Her with a magnetic field strength of only 100\,G by \citet{Kochukhov2023} questioned the necessity of strong magnetic fields to stabilize the spots. Furthermore, precise space observations have revealed that CP stars without strong magnetic fields (HgMn and AmFm stars) and normal A-type stars also show rotation modulation \citep[e.g.][]{Balona2011, Sikora2019, Kochukhov2021, Trust2020}. The brightness variations in the non-magnetic A-stars are less regular than in magnetic CP stars and resemble differential rotation and spot evolution observed in cool stars \citep[e.g.][]{Balona2011, Blazere2020}. This observational evidence, combined with the spot evolution in some HgMn stars \citep{Kochukhov2007} and rotation periods of less than 1 day observed in some Am stars \citep{Trust2020}, challenges the notion of requirements for stable and calm atmospheres in these stars.\footnote{It is worth to note that the short period may arise from binarity that has not been addressed in \citet{Trust2020}.}

The Fourier spectrum of the data series of many normal and AmFm stars displays a broad group of close-spaced peaks, often containing a single peak or a very narrow group of unresolved peaks \citep{Balona2013, Balona2015, Trust2020, Henriksen2023}. 
It has been only recently shown that the sharp peak (referred to as the `spike') is due to surface rotational modulation connected with stellar spots and complex magnetic fields generated in the subsurface convective layer \citep{Antoci2025}.
The broad group of peaks (the `hump') is thought to be either due to prograde g-modes (spike at lower frequency) or unresolved Rossby modes (spike at higher frequency). The Rossby modes are mechanically excited by deviated flows caused by stellar spots or mass outbursts, and by non-synchronous tidal forces \citep{Saio2018rosby}. 

Our study focuses on a 5.3-mag star 50\,Dra (basic parameters in Table~\ref{Tab:Star}), which is a double-line spectroscopic binary system. The binary nature of 50\,Dra was first discovered by \citet{Harper1919}, who found an orbital period of 4.1175~days and estimated the basic parameters of the orbit. \citet{Skarka2022} classified this star as a ROTM|GDOR variable, suggesting variations connected with rotation and/or pulsations. We collected new spectroscopic observations over a century later and found almost the same orbital parameters as \citet{Harper1919}. However, a combination of our new spectroscopic observations with photometric data from the TESS mission \citep[Sect.~\ref{Sect:Data},][]{Ricker2015} allowed us to discover ellipsoidal and additional brightness variations (Sect.~\ref{Sect:PhotVariability}). This enabled us to determine the parameters of the system and both components (Sect.~\ref{Sect:Binary}), and to reveal that both stars are metallic-line CP stars (Sect.~\ref{Sect:SpectralSynthesis}). All the features of 50\,Dra are discussed in Sect.~\ref{Sect:Discussion}.  

\begin{table}
\caption{Basic characteristics of 50\,Dra.}       
\label{Tab:Star}      
\centering     
\begin{tabular}{l l l}       
\hline\hline   
\multicolumn{3}{c}{HD\,175286, TIC 424391564, } \\ \multicolumn{3}{c}{Gaia DR3 2268467486545969792}\\ \hline
ID & Value & Source \\ \hline
RA$_{\rm J2000}$   ($^{\rm hh}:^{\rm mm}:^{\rm ss}$) & 18:46:22.24  & 1 \\
DEC$_{\rm J2000}$ ($^{\circ}:^{\rm '}:^{\rm ''}$) & +75:26:02.24  & 1 \\
Tycho $V_{T}$ (mag) & 5.358(1)& 2\\
Tycho $B_{T}$ (mag) & 5.409(14)& 2\\
TESS $T$ (mag) & 5.345 (7)& 3\\
Gaia $G$ (mag) & 5.357(3)& 4\\
$\mu_{\alpha}\cos\delta$ (mas\,yr$^{-1}$) & 17.06(14) & 4\\
$\mu_{\delta}$ (mas\,yr$^{-1}$) & 70.39(15)& 4\\
Parallax (mas) & 11.42(11)& 4\\
$\gamma$ (km\,s$^{-1}$) & -8.79(49) & 5 \\
    & -7.8 & 6 \\
    & -8.8(2.8) & 7 \\
$T_{\rm eff}$ (K) & 9150(142) & 3\\ 
    & 9572$^{+128}_{-297}$ & 4\\
    & 9130$^{+290}_{-262}$ & SED \\
$\rm{[Fe/H]}$ (dex) & 0.22$^{+0.21}_{-0.14}$ & 4\\
    & -0.07$^{+0.21}_{-0.23}$ & SED\\
$\log g$ (cm\,s$^{-2}$) & 3.97(67) & 3 \\
    & 3.935$^{+0.024}_{-0.028}$ & 4 \\
    & 3.90$^{+0.33}_{-0.33}$ & SED \\
\hline \hline
\end{tabular}
\tablefoot{Note: effective temperature $T_{\rm eff}$ and surface gravity $\log g$ corresponds with the assumption of a single star. \textbf{References:} 1 -- \citet{GaiaEDR3}, 2 -- \citet{Hog2000}, 3 -- \citet{Paegert2021}, 4 -- \citet{GaiaDR3}, 5 -- \citet{Harper1919}, 6 -- \citet{Wilson1953}, 7 -- \citet{Gotcharov2006}, SED -- this work (spectra energy distribution fitting).}
\end{table}

\section{Observations}\label{Sect:Data}
\subsection{TESS photometry}

We collected available data reduced by the TESS Science Processing Operations Center \citep[SPOC;][]{Jenkins2016} and the quick-look pipeline \citep[QLP;][]{Huang2020a, Huang2020b} using \textsc{Lightkurve} software \citep{Lightkurve2018, Barentsen2020} from the MAST archive. We extracted the pre-search data conditioning simple aperture photometry (PDCSAP) flux with long-term trends removed \citep[][]{Twicken2010} and transformed the normalised flux to magnitudes. The \textsc{Lightkurve} was also used for stitching the data from different sectors together. 

\begin{figure}
\centering
\includegraphics[width=0.48\textwidth]{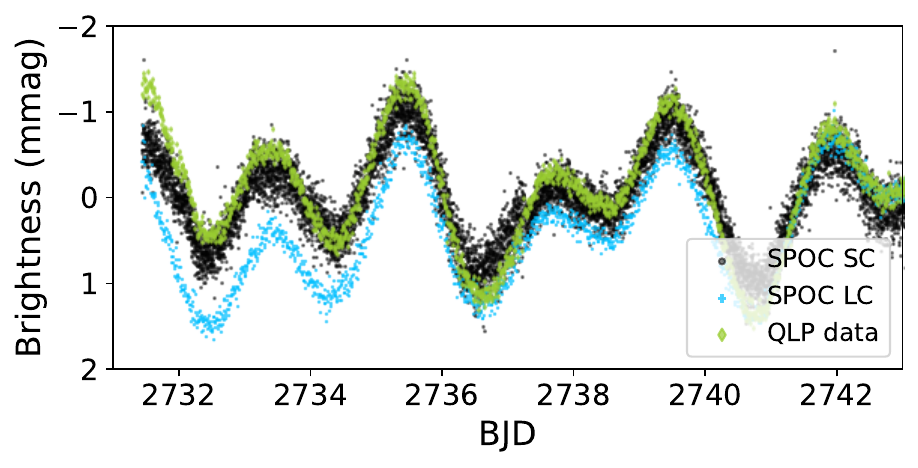}\\
\includegraphics[width=0.48\textwidth]{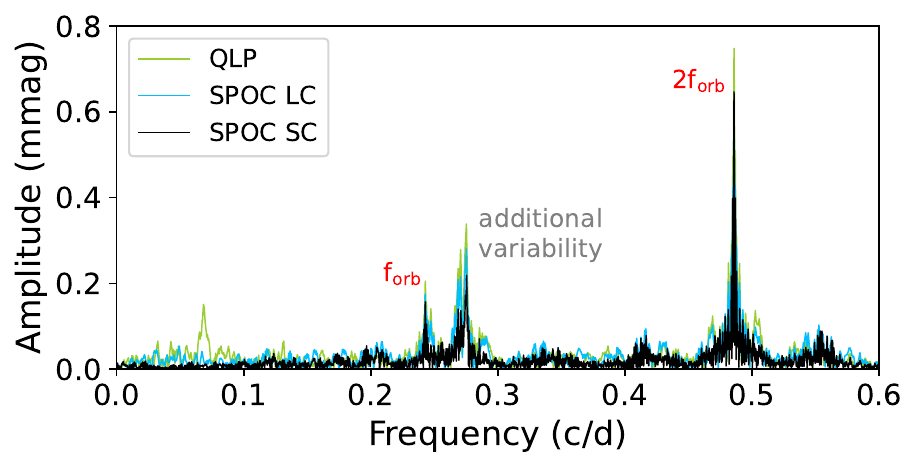}
\caption{Comparison of the available data products produced by different pipelines (top panel) and corresponding frequency spectra with labelled features (bottom panel).
}
\label{Fig:RoutinesComp}
\end{figure}

Data generated by various pipelines at different cadences exhibit differences. The most reliable products are 2-min (short-cadence, SC) SPOC data sets as discussed by \citet{Skarka2022}. The distinctions between the 50\,Dra data products are illustrated in Fig.~\ref{Fig:RoutinesComp}. The frequency spectra of the SPOC SC data exhibit the lowest noise level, do not show the artificial peak at 0.07\,c/d (as is present in QLP), and the distribution of SC data diminishes the presence of the artificial data peaks, for example, around the dominant frequency peak at 0.48\,c/d. Consequently, we opted to base our analysis on the 2-minute SPOC data. We acquired SPOC SC data from 28 sectors (14-26, 40-41, 47-58, 60, and 74), excluding data from sector 25 due to its poor quality. In total, we utilised 440\,166 2-min cadence observations spanning almost 4.5\,years (1629\,days, from 2019-2024).

\begin{figure}
\centering
\includegraphics[width=0.48\textwidth]{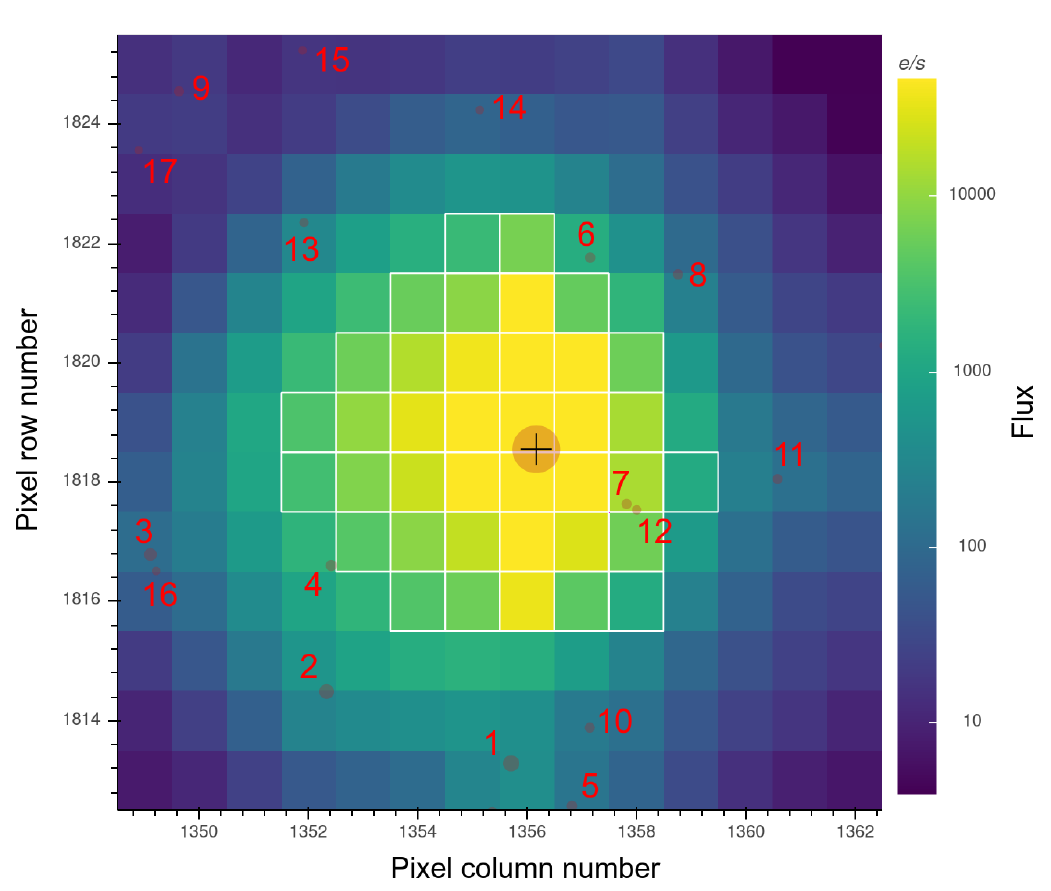}
\caption{Vicinity of 50\,Dra showing the aperture mask in Sector 14 with the identification of possible contaminants identified in Table~\ref{Tab:NearbyStars} in Appendix. The field size shown in the figure is about 4x4\,arcmin.
}
\label{Fig:Mask}
\end{figure}

The contamination ratio of only 0.02\,\% \citep{Paegert2021} suggests no contamination of the 50\,Dra light. The only possible contaminants are two bright stars 20 and 34\,arcmin away\footnote{HD\,174257 ($V=7.53$\,mag) and HD\,176795 ($V=6.71$\,mag)} and 17 additional faint stars (7.7-12.5-mag fainter than 50\,Dra, see Table~\ref{Tab:NearbyStars}) near 50\,Dra shown in Fig.~\ref{Fig:Mask}. However, the two bright stars do not show signatures of variability similar to 50\,Dra and a custom aperture analysis around the faint numbered stars in Fig.~\ref{Fig:Mask} rejects the possibility that any of the signals observed in 50\,Dra originates from a different star. 

\subsection{Spectroscopy}\label{Subsect:Spectroscopy}

We obtained 20 spectra of 50\,Dra between February and July 2022 using the Ond\v{r}ejov Echelle Spectrograph (OES) at the 2m Perek telescope (Ond\v{r}ejov, Czech Republic). The spectrograph has a resolving power of $R=\lambda/\delta \lambda\approx$50\,000 in the H$\alpha$ region and covers a spectral range of 3800-9100\,\AA\ \citep{Koubsky2004, Kabath2020}. The spectra were processed and reduced using standard tasks in the \textsc{Iraf} package \citep{Tody1986}, and the cosmic-particle hits were eliminated using the \textsc{DCR} code \citep{Pych2004}. The median S/N of the 600-second exposures in the H$\alpha$ region was 150, with only four exposures having S/N slightly less than 100, and a few reaching S/N$=230$. 

\section{Photometric variability}\label{Sect:PhotVariability}

We observed the expected double-wave variations in the TESS SC light curve, with a period of $P_{\mathrm{orb}}=4.117719(2)$\,days (peaks labelled $f_{\rm orb}$ in Fig.~\ref{Fig:RoutinesComp}). This period agrees with the orbital period derived from spectroscopic observations but is more precise due to a larger time span. As expected, a primary minimum occurs at the inferior conjunction of the binary components. 

\begin{figure}
\centering
\includegraphics[width=0.49\textwidth]{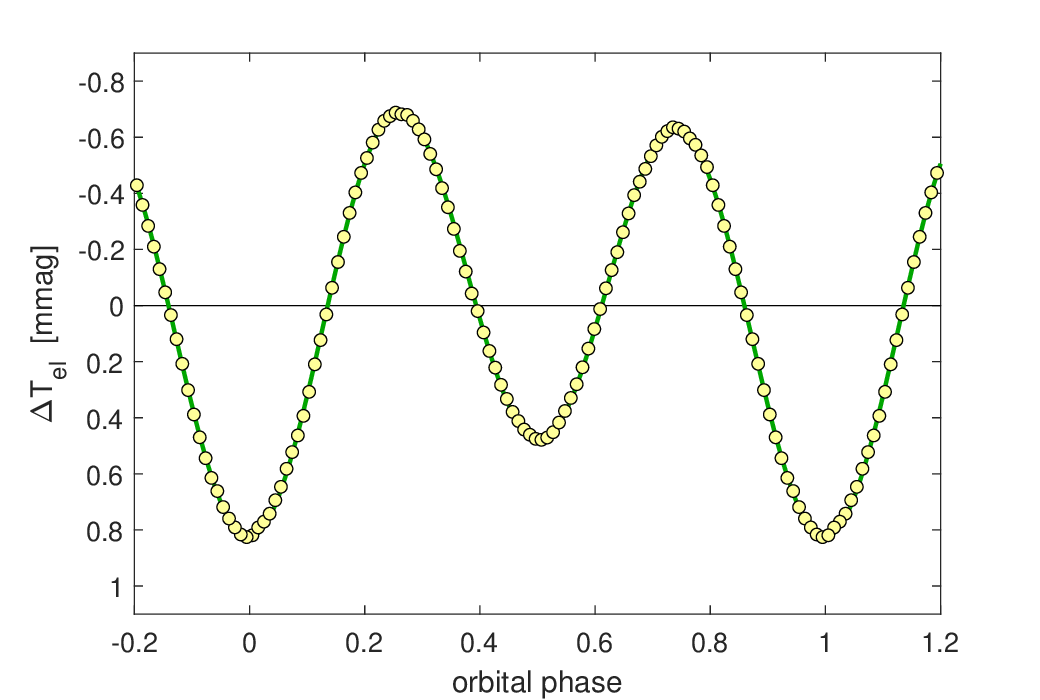}
\caption{TESS SC data of 50\,Dra phase-folded with the orbital period. Each circle corresponds to a mean of 4\,400 individual TESS observations.}
\label{Fig:LCell}
\end{figure}

Given that the orbital period is relatively short and the radial velocity (RV) curves of the components appear almost perfect sinusoids within the observational uncertainties, it is reasonable to assume that the trajectory of the components of the binary star is nearly circular (see Sect.~\ref{Sect:Binary}, Table~\ref{Tab:SystemParameters} and Fig.\,\ref{Fig:RVs}). As a result, the light curve of this non-eclipsing binary, $F_{\mathrm{ell}(t)}$, can be well approximated by a simple trigonometric polynomial model: 
\begin{equation}
   F_{\mathrm{ell}}(t)=\overline{m}+\sum_{k=1}^3 A_k\cos(2\pi k \vartheta)-A_4\sin(2\pi\vartheta),\ \ \vartheta=\frac{t-M_0}{P_{\mathrm{orb}}}, \label{Eq:ellipse}
\end{equation}
 where $t$ is the BJD timestamp, $M_0$ is the reference time for the start of the phase that we set at the moment of the inferior conjunction (the time of the deeper minimum). The phase function ($\vartheta(t)$) is the sum of the number of orbits completed by the system since the passage through the basic inferior conjunction, and the fractional orbital phase $\varphi=\mathrm{Frac}(\vartheta)$. The coefficients of the model are denoted as $A_1,\,A_2,\,A_3,\,A_4$, and $\overline{m}$. The figure displaying the binned data along with the model is shown in Fig.~\ref{Fig:LCell}.

Symmetrical terms of the ellipsoidal variability correspond to the effects of tidal deformation of components and reflection, while the anti-symmetrical term, causing the uneven heights of maxima, is the consequence of the Doppler beaming \citep[e.g.][]{Zucker2007}. Parameters of the model with Eq.~\ref{Eq:ellipse} are in Table\,\ref{Tab:ellpar}. 
 
\begin{table}
\caption{Model parameters of the ellipsoidal variations.}
\label{Tab:ellpar}      
\centering 
\begin{tabular}{l  l}     
\hline\hline
Ephemeris&Parameters\\
\hline
$M_0=2\,459\,365.9508(2)$ & $A_1=0.172\,8(5)$ mmag\\
$P_{\mathrm{orb}}=4\fd117\,719(2)$ & $A_2=0.652\,9(4)$ mmag\\
 & $A_3=0.002\,9(4)$ mmag\\
ampl$_{\mathrm{eff}}=1.35$ mmag&$A_4=0.029\,6(5)$ mmag\\
\hline \hline  
\end{tabular}
\end{table}

\begin{figure}
\centering
\includegraphics[width=0.49\textwidth]{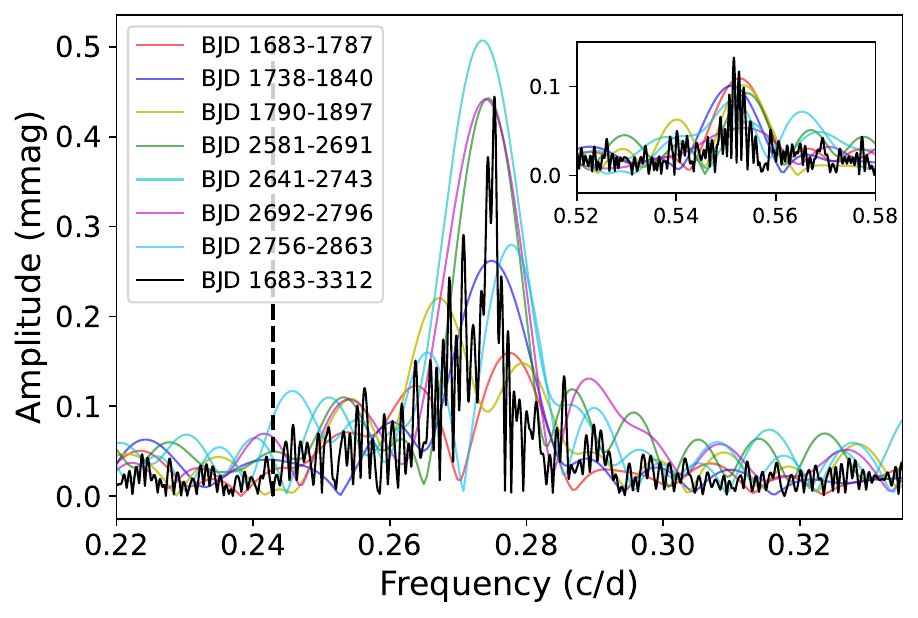}
\caption{Frequency spectrum centered around a group of peaks produced by additional variability from different data segments (colour lines) and the full data set (black line). The amplitude of the Frequency spectrum of the full dataset is multiplied by two for better readability. The black dashed line shows the position of the orbital/rotational frequency.}
\label{Fig:AddVar}
\end{figure}

After removing the variability linked to the orbital period $P_{\mathrm{orb}}=4.117719$\,days, a complex, unresolved variability around 0.27 and 0.55\,c/d remains in the frequency spectrum (see Fig.~\ref{Fig:RoutinesComp} and \ref{Fig:AddVar}). Since the group of peaks with higher frequencies are in the region of harmonics and combination peaks of the lower frequencies, we assume that the peaks have a common nature. As can be seen from Fig.~\ref{Fig:AddVar}, which shows the frequency spectra of approximately 100-days segments, the variation changes over time. We discuss possible explanations of this additional variability in Sect.~\ref{Sect:Discussion}. Apart from the peaks caused by the ellipsoidal variability and the unresolved peaks around 0.27 and 0.55\,c/d, we did not find any other significant frequencies. 



\section{System parameters}\label{Sect:Binary}

\subsection{Spectral energy distribution}\label{Sect:SED}
We used photometry in 12 photometric filters in visual to infra-red wavelengths (Table~\ref{Tab:SED} in Appendix) and fitted the spectral-energy distribution (SED) with \textsc{Ariadne} \citep{Ariadne2022}. This code uses SED fitting methods, and Gaia distances, and combines results via Bayesian model averaging to derive basic stellar parameters such as $T_{\rm eff}$, $\log g$, iron abundance [Fe/H], extinction, and radius of a star.

\begin{figure}
\centering
\includegraphics[width=0.48\textwidth]{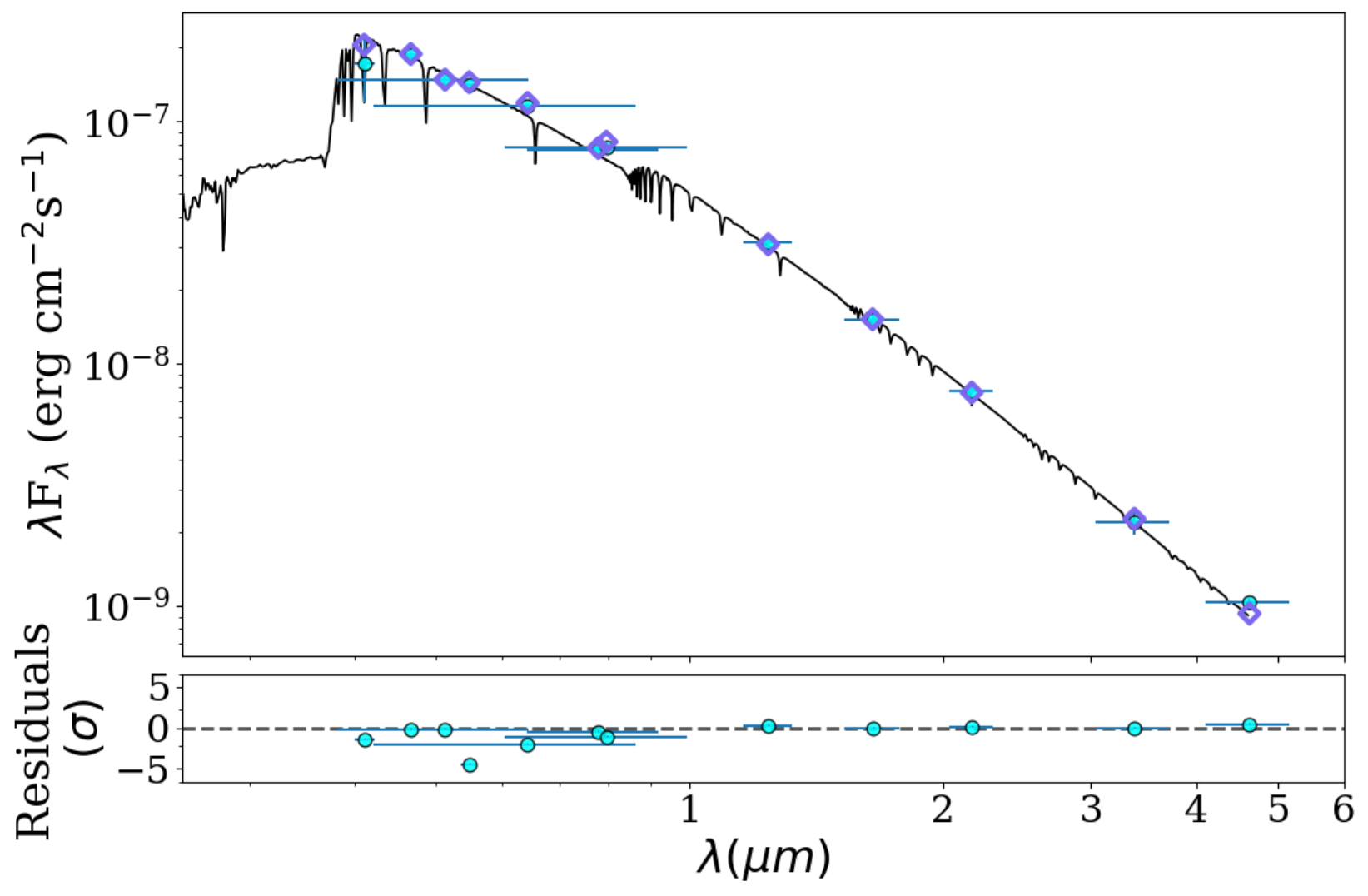}
\caption{Spectral-energy-distribution function based on the photometric observations (Table~\ref{Tab:SED}) from \textsc{Ariadne}.
}
\label{Fig:SED}
\end{figure}

The results of the SED fitting are shown in Fig.~\ref{Fig:SED} and the values we obtained are shown in Table~\ref{Tab:Star} (label `SED' in the column `Source'). The results are given only for reference since 50\,Dra is not a single star. However, this exercise gives a good idea about the reliability of the stellar parameters from the literature when considering 50\,Dra as a single object. The temperature \tef$=9123$\,K, [Fe/H]$=-0.07$\,dex and $\log g=3.90$ are within their errors consistent with the catalogue values, especially \tef~is in excellent agreement with the value from \citet{Paegert2021}. The biggest difference is in the iron abundance which is about 0.3\,dex lower than [Fe/H] from \citet{GaiaDR3} and from our spectroscopic analysis (see Sect.~\ref{Tab:SpectralAnalysis}). Thus, the [Fe/H] form SED is less reliable than from other methods. We will use the radius $R_*=2.91(9)\, R_\odot$ derived using \textsc{Ariadne} under the single star assumption to derive the radii of the components in Sect.~\ref{Subsect:BinaryModel}. 
This is possible because, as shown from the spectral synthesis (see Sect.~\ref{Sect:SpectralSynthesis}), the stars have similar temperatures in range $9000-10000\,\mathrm{K}$ causing the SEDs of individual stars to be almost scaled-up versions of each other in the visual and infrared bands used to derive the $R_*$ with a better approximation in longer wavelengths. We note that binary SED fit alone would not converge to a well-defined solution even using the priors on temperature from spectra because of the degeneracy between the component radii. In case of equal temperatures we would get only a constraint $R_*^2=R_1^2+R_2^2$. 

\subsection{Spectra disentanglement}\label{Subsect:Disentanglement}

\begin{figure}
\centering
\includegraphics[width=0.48\textwidth]{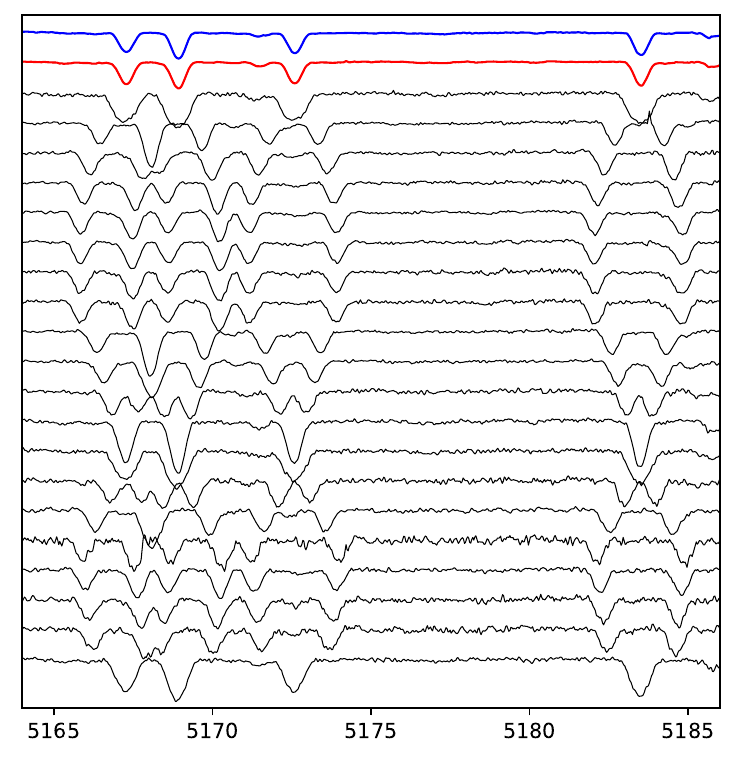}\\
\includegraphics[width=0.16\textwidth]{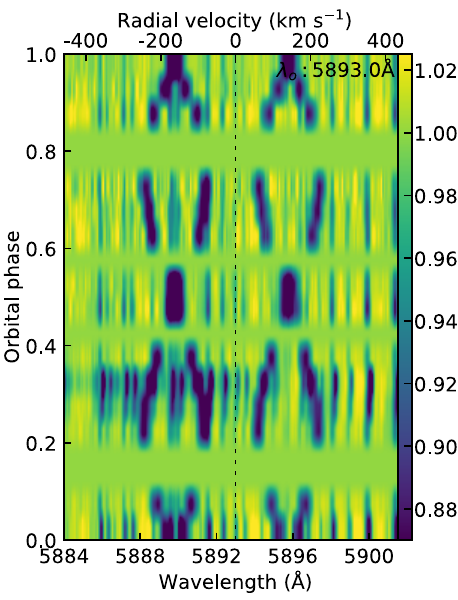} 
\includegraphics[width=0.16\textwidth]{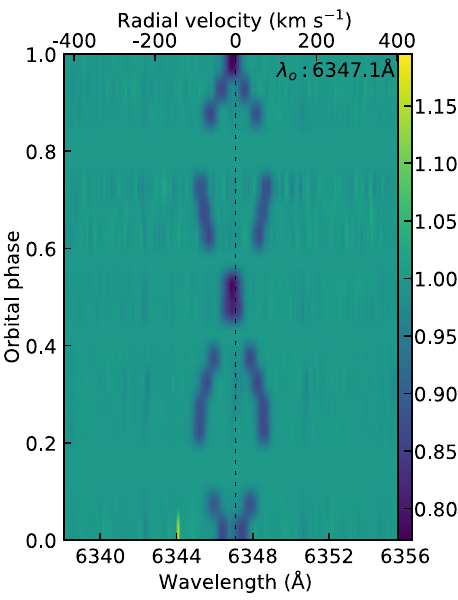}
\includegraphics[width=0.16\textwidth]{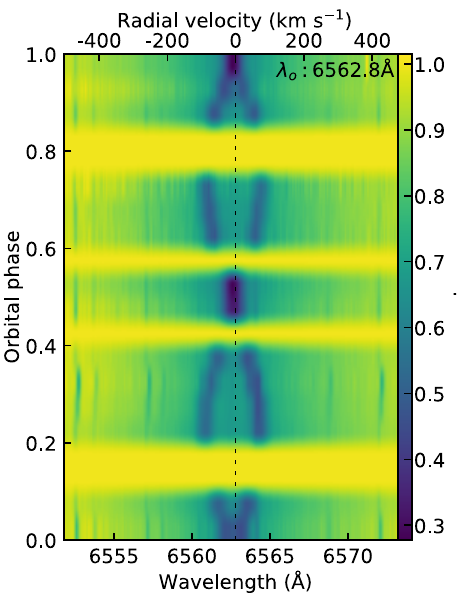}
\caption{{\it (Top)} Observed spectra in different orbital phases around the Mg~I 5167-5183\,\AA\ triplet and Fe~II 5169\,\AA\ lines. The two upper-most spectra show the mean disentangled spectra for the primary (blue) and the secondary component (red). {\it(Bottom)} Trails of Na~D, Si~II, and H$\alpha$ lines (from left to right) showing the variation of the position of the lines of both components during the orbital cycle. The relative intensity of the lines is also indicated.
}
\label{Fig:Spectra}
\end{figure}

Since 50\,Dra is an SB2 binary system, the variations in the position of the spectral lines of both components are distinct and apparent at first glance (see Fig.~\ref{Fig:Spectra}). The movement of some of the prominent spectral lines during the orbital motion is best seen from the trail plots shown in the bottom part of Fig.~\ref{Fig:Spectra}. We used the \textsc{Korel} code \citep[][]{Hadrava1995,Hadrava2004} for the Fourier spectral disentangling. This code performs simultaneous decomposition of spectra and solution of orbital parameters. We fixed the orbital period $P=4.117719$\,days since it has been precisely determined from the ellipsoidal variations detected in the photometric observations with a time span of almost 4.5\,years (Sect.~\ref{Sect:PhotVariability}).

We performed the Fourier spectral disentangling in 41 spectral regions with a typical width of 140\,\AA. An example of the final disentangled spectra of both components around the Mg~I 5167-5183\,$\AA$ triplet is shown in detail in the top part of the top panel of Fig.~\ref{Fig:Spectra}.  \textsc{Korel} also produces RVs and provides a model of the orbit (see Sect.~\ref{Subsect:OrbitalParameters} and Table~\ref{Tab:SystemParameters}). In the end, we decided to use only 25 spectral regions with a sufficient number of spectral lines to get results consistent with each other. The RV values for both components are in Table~\ref{Tab:RVs} together with their errors calculated as the standard deviation of the values from the individual segments. 

\begin{table}
\caption{Radial velocities of both components in km\,s$^{-1}$.}       
\label{Tab:RVs}      
\centering     
\begin{tabular}{c r c r c}       
\hline\hline   
BJD	&	$RV_{1}$	&	$\sigma_{RV_{1}}$	&	$RV_{2}$	&	$\sigma_{RV_{2}}$	\\
2459623.2165	&	-10.95	&	0.65	&	10.45	&	0.57	\\
2459624.2168	&	77.53	&	0.48	&	-81.65	&	0.45	\\
2459625.2641	&	11.68	&	0.67	&	-12.41	&	0.54	\\
2459630.2691	&	-73.43	&	0.34	&	77.75	&	0.35	\\
2459638.3844	&	-67.18	&	0.38	&	71.13	&	0.38	\\
2459639.5434	&	-26.94	&	0.51	&	28.70	&	0.46	\\
2459640.6760	&	77.22	&	0.36	&	-81.54	&	0.41	\\
2459641.4470	&	44.24	&	0.55	&	-46.18	&	0.65	\\
2459646.5317	&	-60.81	&	0.67	&	64.68	&	0.46	\\
2459647.4978	&	-55.29	&	0.46	&	58.77	&	0.43	\\
2459648.4501	&	48.68	&	1.35	&	-51.43	&	0.67	\\
2459652.4616	&	38.66	&	0.71	&	-39.85	&	0.84	\\
2459653.4062	&	73.58	&	0.38	&	-76.96	&	0.41	\\
2459658.3811	&	-9.75	&	0.67	&	9.12	&	0.60	\\
2459659.3145	&	-78.72	&	0.33	&	83.21	&	0.39	\\
2459660.3422	&	-4.29	&	1.62	&	2.26	&	0.56	\\
2459661.3819	&	78.90	&	0.23	&	-82.76	&	0.34	\\
2459722.3264	&	23.24	&	0.47	&	-24.11	&	0.72	\\
2459764.4223	&	78.18	&	0.68	&	-82.31	&	0.62	\\
2459789.4843	&	61.93	&	0.95	&	-64.85	&	0.51	\\
\hline \hline
\end{tabular}
\end{table}

\subsection{Orbital parameters}\label{Subsect:OrbitalParameters}

\begin{figure}
\centering
\includegraphics[width=0.48\textwidth]{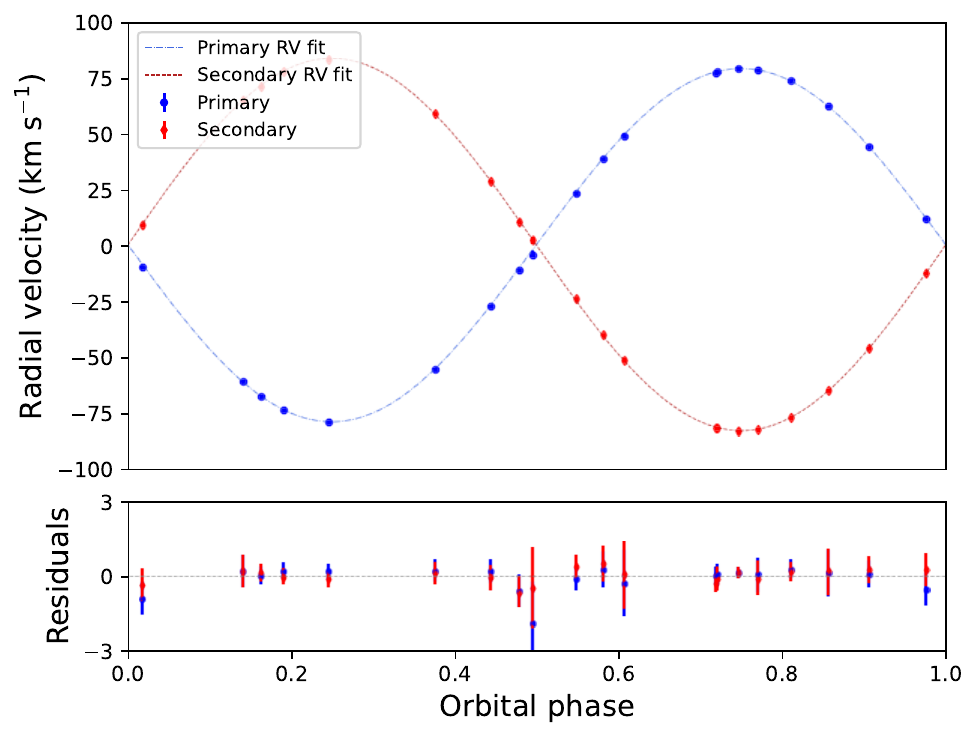}\\
\includegraphics[width=0.48\textwidth]{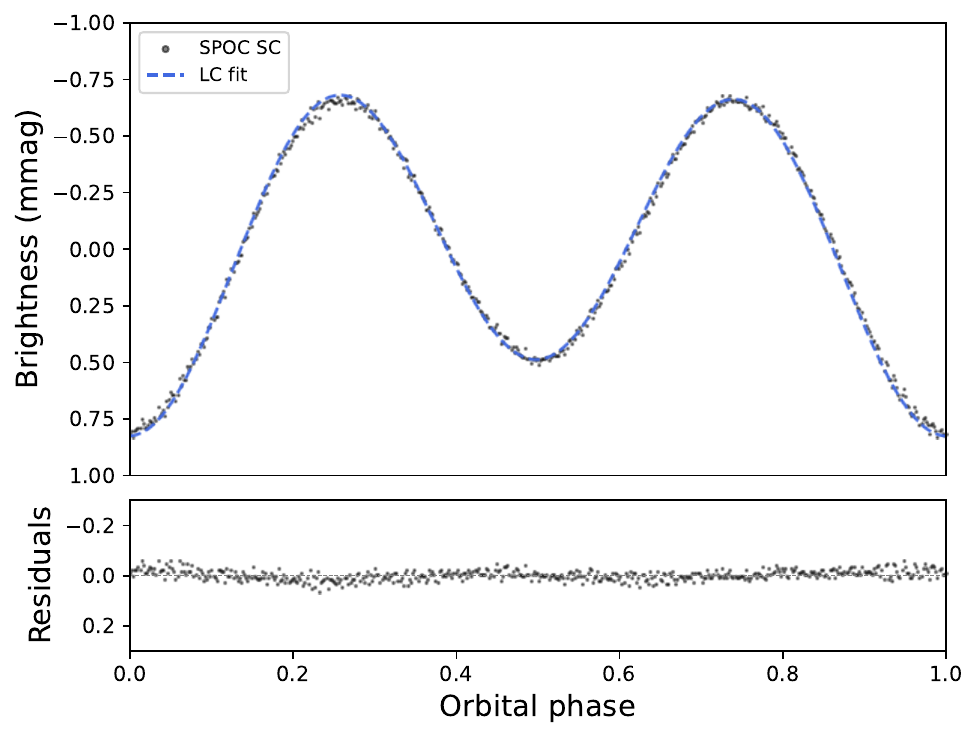}
\caption{Radial velocity curve of both components with the best fit (upper panel) and the light curve with the best fit (bottom panel). The photometric points are binned to have 500 points per orbital phase.
}
\label{Fig:RVs}
\end{figure}

The estimation of the orbital parameters (time of periastron passage $T_{0}$, eccentricity $e$, argument of pericentre $\omega$, semi-amplitudes of radial velocities $K_{1}$, $K_{2}$, and mass ratio $q=M_{2}/M_{1}$) comes from the disentangling process with \textsc{Korel} when the orbit is simultaneously solved for all the disentangled regions. The values in Table~\ref{Tab:SystemParameters}, which are calculated as the average values from all the solutions in 25 spectral regions, signalize circular orbit ($e$ is almost zero) with large semi-amplitudes of the radial-velocity (RV) curves ($K_{1}\approx79$\,km\,s$^{-1}$, $K_{2}\approx83$\,km\,s$^{-1}$). The similarity of the semi-amplitudes and, thus, the mass ratio $q\approx 0.95$, show that both components are almost of equal masses (see Fig.~\ref{Fig:RVs}). Systems with $q\simeq 1$ components are commonly observed among AmFm binaries as was observed, for example, by \citet{Carquillat2007} who found that six of twelve of their SB2 systems have $q>0.9$. It is worth noting that all our values are consistent with what was estimated more than a hundred years ago by \cite{Harper1919}. 

    \begin{table*}
    \caption{Parameters of the binary system.}       
    \label{Tab:SystemParameters}      
    \centering     
    \begin{tabular}{l l l | l l l}       
    \hline\hline   
    	&	\textsc{Korel}			 & H19 & &	\textsc{Param 1.5}	& \textsc{Phoebe 2.4} \\ \hline
    $T_{0}$ &	2459654.241(4)  &  & $M_{1}$\,(\msun) &	2.39$^{+0.27}_{-0.15}$ &	2.08(8)\\
    e	             &	0.0021(3)	  & 0.012(9) & $R_{1}$\,(\rsun) & 2.13$^{+0.80}_{-0.32}$& 2.06(9)\\
    $\omega$ (deg)	 &	94.85(2)     &  & $(\log g)_{1}$ (cm\,s$^{-2}$)	& 4.16$^{+0.23}_{-0.11}$ & 4.13(5)	\\
    $K_{1}$ (km\,s$^{-1}$)	 &	78.93(2)    & 79.12(97)& $M_{2}$\,(\msun) &	2.21$^{+0.15}_{-0.13}$ &	1.97(8)	\\
    $K_{2}$ (km\,s$^{-1}$)	 &	82.96(20)   & 83.9(97) & $R_{2}$\,(\rsun) & 2.34$^{+0.35}_{-0.30}$ & 1.99(9)\\
    $q(M_{2}/M_{1})$ &	0.951(2)    & 0.947 & $(\log g)_{2}$ (cm\,s$^{-2}$)	& 4.04$^{+0.10}_{-0.10}$ & 	4.13(5)\\
    $a\sin i$ (\rsun) &       & 13.3 & $a\sin i$ (\rsun) & & 13.184(26) \\
    $i$ (deg) & & & $i$ (deg) & & 49.9(8) \\
    \hline \hline
    \end{tabular}
    \tablefoot{H19 is the reference to \citet{Harper1919}.}
    \end{table*}

The only discrepancy between our results and literature values is the systemic velocity $\gamma$. Our value $\gamma=-6.53(9)$\,km\,s$^{-1}$ is about 1-2\,km\,s$^{-1}$ larger than the literature values (see Table~\ref{Tab:Star}). We did not identify any problem in our analysis that can explain the difference. Since all the previous values are based on low-resolution spectroscopy and/or on photographic plates taken a long time ago, we assume that our new value is of better reliability than those published previously.


\subsection{Characteristics of the system components}\label{Subsect:BinaryModel}

As a proxy, we estimated stellar parameters using the Bayesian fitting tool \textsc{PARAM 1.5}\footnote{\url{http://stev.oapd.inaf.it/cgi-bin/param}} \citep{daSilva2006, Rodrigues2014, Rodrigues2017}. The code uses a new version of PARSEC \citep{Bressan2012} evolutionary tracks and isochrones that include effects of rotation, improvements in nuclear reaction network and other effects \citep{Nguyen2022}. Since 50\,Dra is not a single star, we had to calculate the observed magnitudes for each component. 

If we assume the mass ratio $q\approx 0.951$ and the mass-luminosity relation $L\simeq M^{4.329}$ for the main-sequence stars in the 1.05--2.4\msun~range \citep[based on 275 well measured stars,][]{Eker2018}, we get the flux ratio $F_{\rm Prim}/F_{\rm Sec}=0.951^{-4.329}=1.243$. After transformation into magnitudes, we end up with $V_{\rm Prim}=5.997$\,mag and $V_{\rm Sec}=6.238$\,mag by assuming the total magnitude of the system $V_{T}=5.358$\,mag \citep{Hog2000}. We used the calculated $V$ magnitudes, \tef, \lgg\ and [Fe/H] from Table~\ref{Tab:SpectralAnalysis}, and Gaia DR3 parallax from Table~\ref{Tab:Star} as the input parameters for \textsc{Param 1.5}. Although being only rough estimates, the resulting values (see Table~\ref{Tab:SystemParameters}) are all within errors consistent with results from binary star modelling and spectral analysis (Sect.~\ref{Sect:SpectralSynthesis}).


As an alternative and more appropriate way of deriving the system parameters without using stellar evolution models, we use the light-curve model, SED fit, and radial velocities. We use \textsc{Phoebe 2.4} \citep{Conroy2020} to run the binary model. Since the Doppler beaming, apparent in our light curve, is not currently implemented in \textsc{Phoebe 2.4} and the use of SED is quite convoluted, we turn to a direct computation of radii and temperatures. 

The Doppler beaming amplitude $\Delta F$ is given by
\begin{equation}
\frac{\Delta F}{F}=\frac{1}{c}\frac{\beta_1 K_1 F_1-\beta_2 K_2 F_2}{F_1+F_2}\,,
\end{equation}
where $c$ is the speed of light and $\beta_1$, $\beta_2$ are Doppler beaming coefficients of the stars. The relative beaming amplitude $\frac{\Delta F}{F}$ taken with respect to the total flux of the binary components $F=F_1+F_2$ can be easily converted from magnitude value $A_4$ in Table\,\ref{Tab:ellpar} corresponding to the expansion in Eq.~\ref{Eq:ellipse}.
Using the radial velocity amplitudes $K_1$, $K_2$ from Section \ref{Subsect:OrbitalParameters} and beaming coefficients $\beta_1=2.01(5)$, $\beta_2=2.09(5)$ derived by interpolation from tables by \citet{2020A&A...641A.157C} for TESS passband we get the passband relative fluxes $F_1/(F_1+F_2)=0.547(9)$ and $F_2/(F_1+F_2)=0.453(9)$. Next, we use \textsc{Phoebe} to derive the relation between the flux and temperature ratios for the stars with the same radius $R_1=R_2=2R_\odot$ in the TESS passband. Using the range $T_2/T_1\in (0.90,1.00)$ and primary star temperature $T_1=9800\,K$ we get a linear relation between the two. After adding the factor of the ratio of surface areas we come to the final relation
\begin{equation}
F_2/F_1=\left(-0.88+1.88 \frac{T_2}{T_1}\right) \left(\frac{R_2}{R_1}\right)^2\,.
\end{equation}
To get the radii themselves, we use the SED result for $R_{*}$. In the long wavelength limit (far IR) under the Raighley-Jeans approximation it should hold
\begin{equation}
    R_1^2 T_1 + R_2^2 T_2= R_*^2T_*\,,
\end{equation}
where $R_*$, and $T_*$ are the stellar radius and effective temperature derived in the single star assumption in Sect. \ref{Sect:SED}. Using the effective temperatures of stars $T_1=9800(100)\,K$ and $T_2=9200(200)\,K$ derived from spectra (see Sect.~\ref{Sect:SpectralSynthesis}) we get the radii $R_1=2.06(9)$ and $R_2=1.99(9)$. 

Next, we can fit the ellipsoidal variation in \textsc{PHOEBE} to get the inclination of the system. We use PHOENIX atmospheric models by \citet{aa19058-12} to get the passband luminosities and limb-darkening coefficients. We set bolometric gravity brightening coefficients of both stars $b_1=b_2=1.0$ and bolometric reflection coefficients to $a_1=a_2=1.0$ valid for stars above $T_1=8000\,K$ with radiative atmospheres (see \citet{2003A&A...406..623C}). Sampling the derived radii-temperature distributions we get the inclination of $i=49.9(8)\deg$. This finally allows us to derive the semimajor axis of the system and masses of the components from the radial velocity fit. The best fit (shown in Fig.~\ref{Fig:RVs}) gives values shown in the last column in Table~\ref{Tab:SystemParameters}. All the parameters obtained with \textsc{Phoebe} are in line with the results from other routines and methods.


\section{Spectral synthesis and abundances}\label{Sect:SpectralSynthesis}

Before the analysis of the spectra, we scaled the disentangled spectra of both components assuming the flux ratio of $F_{1}/F_{2}=1.243$ \citep[based on empirical formulae from][]{Eker2018} meaning that $F_{1}=0.555$\,F$_{\rm tot}$ and $F_{2}=0.445$\,F$_{\rm tot}$ which are in agreement with flux ratios calculated from the Doppler beaming amplitudes (Sect.~\ref{Subsect:BinaryModel}). We used the spectrum synthesis method to analyse the spectra of both stars. This method allows for the simultaneous determination of parameters influencing stellar spectra and involves minimising the deviation between theoretical and observed spectra. The synthetic spectrum depends on stellar parameters such as effective temperature (\tef), surface gravity (\lgg), microturbulence (\vmic), projected rotational velocity ($V\sin i$), and the relative abundances of the elements ($\log N(\rm El)$), where `El' denotes the individual element. All these parameters are correlated.

All the necessary atmospheric models were computed with the line-blanketed, local thermodynamical equilibrium (LTE) ATLAS9 code whereas the synthetic spectra were computed with the SYNTHE code \citep{2005MSAIS...8...14K}. Both codes, ATLAS9 and SYNTHE, were ported to GNU/Linux by \cite{2005MSAIS...8...61S}. The stellar line identification and the abundance analysis over the entire observed spectral range were performed based on the line list from the Fiorella Castelli website\footnote{\url{https://wwwuser.oats.inaf.it/fiorella.castelli/}}. The solar abundances were adopted from \citet{2005ASPC..336...25A}.

In our method, effective temperature, surface gravity and microturbulence were obtained from the analysis of lines of neutral and ionised iron. We adjusted \tef, \lgg\ and \vmic\ by comparing the abundances determined from Fe~I and Fe~II lines. First, we adjusted \vmic\ until we saw no correlation between iron abundances and line depths for the Fe~I lines. Next, we changed \tef\ until there was no trend in the abundance versus excitation potential for the Fe~I lines. Then, surface gravity was obtained by fitting the Fe~II and Fe~I lines, ensuring the same iron abundances from the lines of both ions. With the derived \tef, \lgg, and \vmic, the determination of abundances was performed. The final results are presented in the Table~\ref{Tab:SpectralAnalysis}. The errors of chemical abundances given in Table~\ref{Tab:SpectralAnalysis} are standard deviations resulting from the analysis of many spectral lines of a given element or result from the steps in the grid of calculated atmospheric models. 

\begin{table*}
\caption{Parameters of the components and their abundances from the spectral synthesis.}       
\label{Tab:SpectralAnalysis}      
\centering     
\begin{tabular}{c | c c c | c c c | c}       
\hline\hline   
                  &	\multicolumn{3}{c|}{Star 1}		&	\multicolumn{3}{c|}{Star 2}	&	\\ \hline 
          \tef~(K) &	\multicolumn{3}{c|}{9800(100)}	&	\multicolumn{3}{c|}{9200(200)} &	\\
$\log g$ (cm\,s$^{-2}$)	& \multicolumn{3}{c|}{4.1(2)} &   \multicolumn{3}{c|}{4.0(1)} &	\\
  \vmic~(km\,s$^{-1}$) &	\multicolumn{3}{c|}{0.5(3)}	&	\multicolumn{3}{c|}{0.5(3)}	&	\\
  \vsin~(km\,s$^{-1}$)&	\multicolumn{3}{c|}{19(1)}	&	\multicolumn{3}{c|}{19(1)}	&	\\ \hline
El	&	Nr	&	$\log \epsilon$	&	$\sigma$	&	Nr	&	$\log \epsilon$	&	$\sigma$	&	$\log \epsilon$ (Sun)	\\ \hline
    C  &   15  &  8.17 &  0.25  &  15  &  8.12 &  0.24 &  8.43  \\
    N  &    4  &  8.08 &  0.35  &   7  &  8.17 &  0.16 &  7.83  \\
    O  &    8  &  8.52 &  0.18  &   6  &  8.82 &  0.18 &  8.69  \\
   Ne  &    -  &   -   &    -   &   1  &  9.07 &  - &  7.93  \\
   Na  &    4  &  6.88 &  0.22  &   1  &  6.28 &  - &  6.24  \\
   Mg  &    9  &  7.46 &  0.23  &   8  &  7.24 &  0.06 &  7.60  \\
   Al  &    2  &  6.42 &  -  &   3  &  6.38 &  0.11 &  6.45  \\
   Si  &   11  &  7.50 &  0.23  &  15  &  7.55 &  0.26 &  7.51  \\
    P  &    1  &  6.26 &  -  &   2  &  6.36 &  - &  5.41  \\
    S  &    9  &  7.75 &  0.19  &  12  &  7.66 &  0.17 &  7.12  \\
   Ca  &   12  &  6.28 &  0.10  &  13  &  6.03 &  0.14 &  6.34  \\
   Sc  &    5  &  2.51 &  0.11  &   7  &  2.33 &  0.38 &  3.15  \\
   Ti  &   42  &  5.14 &  0.12  &  39  &  4.94 &  0.15 &  4.95  \\
    V  &    8  &  4.43 &  0.06  &   8  &  4.56 &  0.31 &  3.93  \\
   Cr  &   55  &  6.09 &  0.13  &  66  &  5.93 &  0.16 &  5.64  \\
   Mn  &   16  &  5.76 &  0.14  &  24  &  5.65 &  0.12 &  5.43  \\
   Fe  &  207  &  7.71 &  0.09  & 206  &  7.58 &  0.10 &  7.50  \\
   Co  &    1  &  5.43 &  -  &   4  &  5.63 &  0.24 &  4.99  \\
   Ni  &   34  &  6.85 &  0.11  &  44  &  6.69 &  0.14 &  6.22  \\
   Cu  &    2  &  4.89 &  -  &   2  &  4.64 &  - &  4.19  \\
   Zn  &    1  &  5.51 &  -  &   2  &  5.14 &  - &  4.56  \\
   Sr  &    1  &  3.89 &  -  &   2  &  3.47 &  - &  2.87  \\
    Y  &    6  &  3.15 &  0.09  &  11  &  3.07 &  0.15 &  2.21  \\
   Zr  &    5  &  3.68 &  0.19  &  16  &  3.40 &  0.10 &  2.58  \\
   Ba  &    4  &  3.59 &  0.04  &   3  &  3.27 &  - &  2.18  \\
   La  &       &       &        &   1  &  2.37 &  - &  1.10  \\
   Ce  &    1  &  3.17 &  -  &   7  &  2.54 &  0.20 &  1.58  \\
   Nd  &    1  &  2.62 &  -  &   2  &  2.33 &  - &  1.42  \\
   Sm  &       &       &        &   1  &  1.95 &  - &  0.96  \\
\hline \hline
\end{tabular}
\tablefoot{The 'Nr' gives the number of lines used to calculate abundance (including blends), $\log \epsilon$ gives the average abundances (on the scale in which log $\epsilon(\rm{H}) = 12$), $\sigma$ gives the standard deviation in case that the fit was based on more than three lines, $\log \epsilon$ (Sun) gives the solar abundance \citep{Asplund2009}.}
\end{table*}

\begin{figure*}
    \includegraphics[width=0.98\textwidth]{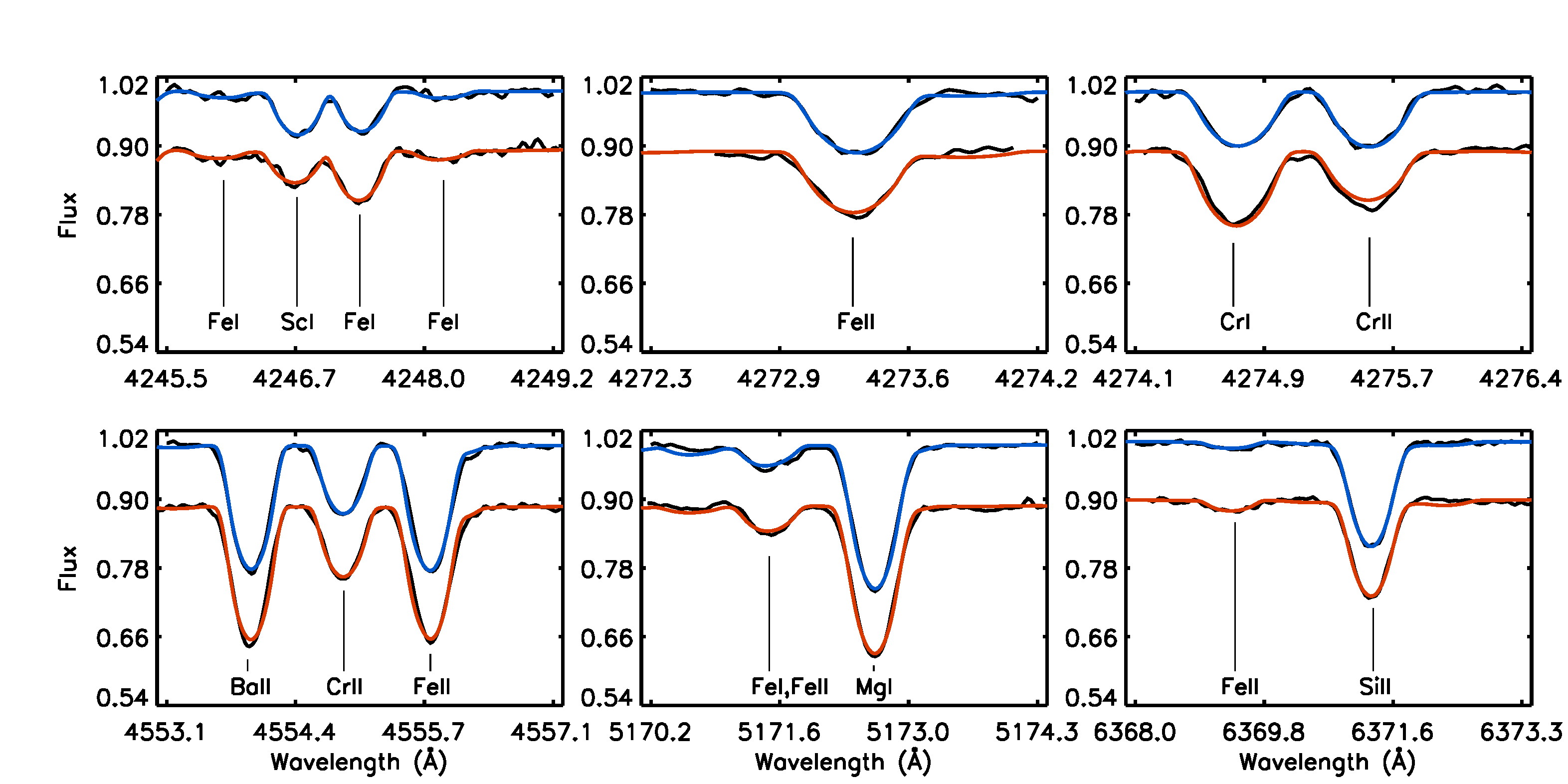}\\
    \caption{Comparison of the observed (black lines) and theoretical spectra calculated for the final parameters. Star 1 is shown with the blue line, and Star 2 with the red line. The spectra of Star 2 were shifted by subtracting a value of 0.1 from the flux.}
\label{Fig:FluxComp}
\end{figure*}

The derived atmospheric parameters and chemical abundances are inﬂuenced by errors from several sources, for example, assumptions taken into account to build an atmospheric model, the adopted atomic data, and spectra normalisation. A detailed discussion of possible uncertainties of the obtained parameters is given in \citet{2015MNRAS.450.2764N} and \citet{Niemczura2017}. The comparison of the observed and theoretical spectra calculated for the final parameters is shown in Fig.~\ref{Fig:FluxComp}. 

The effective temperatures of both components (\tef$_{1}=9800(100)$\,K and \tef$_{2}=9200(200)$\,K) and their surface gravities ((\lgg)$_{1}=4.1(1)$ and (\lgg)$_{2}=4.0(1)$\,cm\,s$^{-2}$) suggest that both stars are of A0-A3~V spectral types. Both system components are slow-rotators with \vsin$=19(1)$\,km\,s$^{-1}$.

We estimated abundances for 29 elements (see Table~\ref{Tab:SpectralAnalysis}). The number of lines used for the abundance estimation is given in columns denoted as `Nr'. It is seen that Fe abundance is the most robust value based on 207 and 206 spectral lines for the primary and secondary components, respectively. The iron abundance is slightly different for both components (${\rm [Fe/H]_{prim}=0.21(9)}$ and ${\rm [Fe/H]_{sec}=0.08(10)}$) but they are consistent within their errors. Both [Fe/H] values are in line with the catalogue value \citep{GaiaDR3} but are about 0.3~dex higher than the values from the SED fitting (Sect.~\ref{Sect:Binary}). The abundances of iron-peak and Earth-rare elements together with slow rotation suggest the chemical peculiarity of both components as discussed in Sect.~\ref{Subsect:ChemPeculiarity}.

\section{Discussion}\label{Sect:Discussion}

\subsection{Rotation}\label{Sect:Rotation}

Slow to moderate rotation is a necessary ingredient for the occurrence of chemical peculiarity. In case of AmFm stars the rotation rate should not exceed about 120\,km\,s$^{-1}$ \citep{Abt1973,Michaud1983}. The rotation velocity of the components $v_{\rm rot, 1,2}=v\sin i / \sin i\approx 24.8$\,km\,s$^{-1}$ (see Table~\ref{Tab:SystemParameters}) is well below this limit and is not special among AmFm stars \citep[see the distribution of rotational velocities in][]{Royer2007,Trust2020,Qin2021,Niemczura2017}. 


If we assume values of $R_{1,2}$, $(v\sin i)_{1,2}$, and $i$ from Table~\ref{Tab:SystemParameters} and their uncertainties, we can calculate rotation frequencies of both components $f_{\rm rot, 1,2}$ and their errors. To do that, we solve a commonly used relation for the rotation period \citep[e.g.][]{Preston1971}
\begin{equation}
    f_{\rm rot} = \frac{v\sin i}{50.6 R_{*}\sin i},
\end{equation}
and use the propagation of errors law. We end up with rotational frequencies of the two components $f_{\rm 1,rot}=0.238(17)$ and $f_{\rm 2,rot}=0.246(17)$\,c/d. Both values are consistent with the orbital frequency $f_{\rm orb}=0.2428529(1)$\,c/d. This simple exercise suggests that the value of the inclination is well established and that the system is relaxed with a circularized orbit and synchronous rotation of the components. The assumption of circular orbit and synchronously rotating components is further reinforced by the following angular momentum investigation. 

The tidal equilibrium of the system is only possible if the total angular momentum of the system $L_{\rm tot}$ (a sum of the orbital and spin momenta) is larger than a critical momentum $L_{\rm crit}$ that can be calculated following \citet{Ogilvie2014} as
\begin{equation}
    L_{\rm crit}=4I(GM)^{1/2}\left(\frac{\mu}{3I}\right)^{3/4},
\end{equation}
where $M=M_{1}+M_{2}$ is the total mass of the stars (values from \textsc{Phoebe} in Table~\ref{Tab:SystemParameters}), $I=I_{1}+I_{2}$ is the total spin momentum of inertia of the stars, and $\mu=M_{1}M_{2}/(M_{1}+M_{2})$. The spin moment of inertia of the stars was calculated as $I=\beta^{2}M_{*}R^{2}_{*}$ with $M_{*}$ and $R_{*}$ from Table~\ref{Tab:SystemParameters} and $\beta=0.218$ for stars with $M=2$\,\msun\ from \citet{Claret1989}.

The total angular momentum of the system $L_{\rm tot}$ 
\begin{equation}
     L_{\rm tot}=L_{\rm orb}+L_{\rm spin}=L_{\rm orb}+ I_{1}\Omega_{1}+I_{2}\Omega_{2}
\end{equation}
is dominated by the orbital angular momentum $L_{\rm orb}$ that can be expressed in a form of
\begin{equation}
    L_{\rm orb}=\frac{GM_{1}M_{2}\left( 1-e^{2}  \right)^{1/2} }{\Omega a},
\end{equation}
where $\Omega=2\pi f_{\rm orb}$\,rad\,s$^{-1}$ is the angular velocity. If we assume that the spin angular velocity of the stars equals the orbital angular velocity (the calculated rotational frequencies match the orbital frequency), the $L_{\rm orb}\approx 380 L_{\rm spin}$. The total angular momentum $L_{\rm tot}\sim 2.5L_{\rm crit}$ means that the system is in tidal equilibrium. At the same time, $L_{\rm orb}\gg 3L_{\rm spin}$ that means that the system is relaxed with a stable tidal equilibrium resulting in circularized orbit and synchronous rotation of the components \citep{Hut1980, Ogilvie2014}. \citet{Carquillat2007} found that the cut-off orbital period for AmFm stars to be circularized is 5.6(5)\,days ($f_{\rm orb}\geq 0.179$\,c/d). All the indices we have pointed towards the synchronous rotation of 50\,Dra components with the orbital frequency 0.243\,c/d. 

\subsection{Additional variability}\label{Sect:Variability}

Since both components are almost equal, it is difficult to assign the additional variability seen as a group of peaks in the frequency spectra (Fig.~\ref{Fig:RoutinesComp} and Fig.~\ref{Fig:AddVar}) to one of the stars and make a proper interpretation. Unresolved pattern in the frequency feature suggests that it results from stochastic semi-regular or unresolved regular variability. In cold stars, such variations can be associated with magnetic activity, cold spots, their finite life span, and differential surface rotation. Although not expected in A-type stars (including AmFm stars), rotational variability resembling the one in cold stars likely connected with spots has been reported in literature \citep[e.g.][]{Balona2011,Sikora2019,Trust2020}.

Stellar spots in cold stars are connected with local magnetic fields. \citet{Cantiello2019} demonstrated that spots in A-type stars can be induced by local magnetic fields generated by convection in thin (sub)surface zones due to high opacity and/or low adiabatic gradient in the ionization zones of H, He, and/or Fe. A weak magnetic field was detected in a few AmFm stars \citep[e.g.][]{Blazere2016a, Blazere2016b}. Particularly, components of 50\,Dra are very similar to Sirius A, which is also an AmFm star, where a very weak magnetic field of the size of $0.2\pm0.1$\,G was detected \citep{Petit2011}, giving a chance that this mechanism could work also in 50\,Dra. However, the spot scenario is less likely because the group of peaks has higher frequency than the rotational frequency (see Sect.~\ref{Sect:Rotation}) and is well separated from it (see Fig.~\ref{Fig:AddVar}). Explanation of such behaviour would require that spots are located exclusively at higher latitudes avoiding areas close to the equator and that the polar regions rotate faster than the equatoreal regions. On the other hand, we still know very little about potential latitudinal differential rotation in the context of binarity. The centre of the power hump is only about 10\,\% higher than the value of $f_{\rm 2,rot}$ giving a chance that differential rotation still may be a plausible explanation that cannot be fully ruled out.

The group of frequencies (hump) could possibly originate from unresolved prograde g-modes that have higher frequency than the rotational frequency. Similar features have been reported by \citet{Saio2018rosby} and \citet{Henriksen2023,Henriksen2023b}. The difference between the hump and the rotational frequency in 50\,Dra is only about 0.03\,c/d while the calculation for a star with two solar masses by \citet{Henriksen2023b} suggests frequency separation of about 0.2\,c/d. However, 50\,Dra is about 2500\,K hotter than their model star and its rotation rate is about four times lower. The explanation of the hump in 50\,Dra via g-modes needs to be proven with detail modelling since g-mode excitation, as seen in $\gamma$\,Dor stars, is not expected an rarely observed at these temperatures \citep{Guzik2000,Dupret2005,Durfeldt-Pedros2024}. 

\subsection{Chemical peculiarity}\label{Subsect:ChemPeculiarity}

To check the chemical peculiarity/normality of 50\,Dra components, we calculated average abundances of CP stars from the catalogue by \citet{Ghazaryan2018} and compared with abundances of both components listed in Table~\ref{Tab:SpectralAnalysis}. The abundances of particular elements in AmFm (118 stars), HgMn (112) and ApBp (188) stars have typical uncertainties of 0.35, 0.65 and 0.60\,dex, respectively. For stars with normal abundances, we used the compilation of 33 stars by \citet{Niemczura2017} with a typical uncertainty of the abundance value of 0.39\,dex. In addition, we plot mean element abundances of 62 Am stars from \citet[][their table~3]{Catanzaro2019}.

\begin{figure*}
\centering
\includegraphics[width=0.96\textwidth]{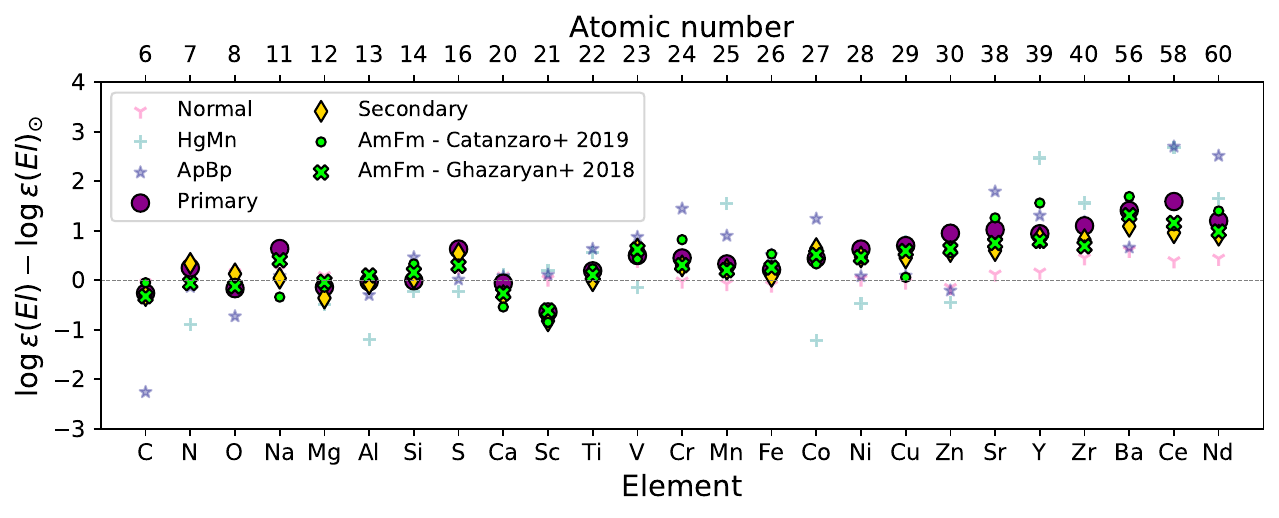}
\caption{Comparison of the mean element abundances of different classes of chemically peculiar stars \citep{Ghazaryan2018, Catanzaro2019}, normal stars \citep{Niemczura2017} and 50\,Dra components (magenta circles, yellow diamonds). The abundances of both stars of 50\,Dra follow the AmFm abundances (green crosses and circles). 
}
\label{Fig:Abundances}
\end{figure*}

By comparing catalogue values with abundances of 50\,Dra components listed in Table~\ref{Tab:SpectralAnalysis}, we see that both stars follow the sequence of AmFm stars (Fig.~\ref{Fig:Abundances}), with low abundance of Sc and overabundant heavy and rare-earth elements. Note that the mean values of AmFm stars from \citet{Ghazaryan2018} are a bit different from those in \citet{Catanzaro2019} increasing the space of possible values for AmFm stars. 

Previous studies have shown that more than 70\,\% of AmFm stars are part binary systems \citep[e.g.][]{Abt1985,Carquillat2007}. In addition, AmFm stars tend to appear in tight binaries with periods less than 20\,days with a peak at about 5\,days \citep{Carquillat2007}. It is believed that the tidal forces in binary systems play a crucial role in slowing down the rotation of these stars, which allows for atomic diffusion and the emergence of the chemical peculiarity observed in AmFm stars. In this context, 50\,Dra is considered a typical representative of this class of CP stars. 

Most of the Am stars have temperatures between 7250 and 8250\,K peaking at 7750\,K \citep{Qin2019}. From their collection of 9372 Am stars, \citet{Qin2019} found only 32 stars to have temperatures higher than 9000\,K and only five to be hotter than 9500\,K. On the other hand, \citet{Catanzaro2019} identified 15 out of their 62 Am sample stars to have temperatures above 9500\,K based on SED fitting. Anyways, components of 50\,Dra belong to a small group of hot Am stars.

\section{Conclusions}\label{Sect:Conclusions}

We performed an analysis of twenty high-resolution spectra from the OES spectrograph \citep{Koubsky2004,Kabath2020} together with photometric data from TESS \citep{Ricker2015} to investigate 50\,Dra. We modelled the radial velocity curve, as well as the photometric data to reveal that the system consists of two intermediate-mass stars with almost the same masses close to 2\,\msun\ ($q=0.951$). The system with an orbital period of 4.117719(2)\,days is observed under an inclination of $i=49.9(8)$\,deg and shows ellipsoidal variations. Our investigation shows that both components are slow rotators with $v\sin i=19(1)$\,km\,s$^{-1}$ that rotate synchronously with the orbital period.

Based on the analysis of separated spectra and comparison with catalogue values, it was determined that both stars in the system are metallic-line AmFm CP stars with temperatures of 9800 and 9200\,K. The high temperatures indicate that components of 50\,Dra belong to a less common group of AmFm stars with temperatures higher than 9000\,K. Apart from the effects of binarity, such as ellipsoidal variation, reflection effect, and beaming, the only feature detected in the frequency spectrum was the hump around 0.275\,c/d. As the most probable explanation of this structure, we assume the presence of prograde g-modes. No signs of p-mode pulsations, that are rare in this temperature region \citep{Durfeldt-Pedros2024}, or other signs of variability were found. We were not able to assign the hump feature to one of the components due to their similarities. 



\begin{acknowledgements}
      MS acknowledges the support by Inter-transfer grant no LTT-20015. PK acknowledges the funding from ESA PRODEX PEA 4000127913. MS is grateful to T. van Reeth for the discussion about the nature of the additional variations and to P. Gajdo\v{s} for reading the manuscript.
      This paper includes data collected with the TESS mission. Funding for the TESS mission is provided by the NASA Explorer Program. Funding for the TESS Asteroseismic Science Operations Centre is provided by the Danish National Research Foundation (Grant agreement no.: DNRF106), ESA PRODEX (PEA 4000119301) and Stellar Astrophysics Centre (SAC) at Aarhus University. We thank the TESS team and staff and TASC/TASOC for their support of the present work. We also thank the TASC WG4 team for their contribution to the selection of targets for 2-minute observations. The TESS data were obtained from the MAST data archive at the Space Telescope Science Institute (STScI). We acknowledge the usage of the data taken with the Perek telescope at the Astronomical Institute of the Czech Academy of Sciences in Ond\v{r}ejov and would like to thank the observers for their work. The calculations have been partly carried out using resources provided by the Wroc\l{}aw Centre for Networking and Supercomputing (http://www.wcss.pl), Grant No 214. This research made use of NASA’s Astrophysics Data System Bibliographic Services, and of the SIMBAD database,
operated at CDS, Strasbourg, France. 
\end{acknowledgements}

\bibliographystyle{aa}
\bibliography{references}

\begin{appendix} \label{Sect:Appendix}
\section{Supporting material}
\begin{table}
\caption{Stars in the vicinity of 50\,Dra shown in Fig.~\ref{Fig:Mask}.}       
\label{Tab:NearbyStars}      
\centering     
\begin{tabular}{l l l l}       
\hline\hline   
NR	&	Gaia DR3 ID	&	G	&	B-R	\\ \hline
1	&	2268466696271988736	&	13.047	&	0.9704	\\
2	&	2268467349107014400	&	13.72	&	0.971	\\
3	&	2268467383466749184	&	14.873	&	0.951	\\
4	&	2268467349107013760	&	15.786	&	1.3158	\\
5	&	2268466627552513024	&	16.034	&	1.0765	\\
6	&	2268467894567507072	&	16.153	&	1.1483	\\
7	&	TIC 335965559	&	16.331	&	-	\\
8	&	2268467898862831360	&	16.407	&	0.9278	\\
9	&	2268467795783608832	&	16.5	&	1.4393	\\
10	&	2268466661912250752	&	16.529	&	1.8639	\\
11	&	2268467074229113344	&	16.703	&	1.0081	\\
12	&	TIC 335965560	&	17.059	&	-	\\
13	&	2268467727064133248	&	17.088	&	1.0404	\\
14	&	2268467928927245952	&	17.624	&	0.8851	\\
15	&	2268467757127995264	&	17.681	&	1.0049	\\
16	&	2268467379171426048	&	17.767	&	1.2799	\\
17	&	2268467688408453248	&	17.856	&	2.0393	\\

\hline \hline
\end{tabular}
\tablefoot{Data from \citet{GaiaDR3}}
\end{table}

\begin{table}
\caption{Magnitudes used for SED fitting.}       
\label{Tab:SED}      
\centering     
\begin{tabular}{l l l l}       
\hline\hline   
Filter	&	Magnitude	&	Uncertainty & Ref\\ \hline
STROMGREN $v$	&	5.598	&	0.013	& 1 \\
STROMGREN $b$	&	5.392	&	0.006	& 1 \\
GaiaDR2v2 {\it BP}	&	5.371	&	0.003 & 2\\
STROMGREN $y$	&	5.369	&	0.001 & 1 \\
GaiaDR2v2 $G$	&	5.357	&	0.003 & 2 \\
GaiaDR2v2 {\it RP}	&	5.303	&	0.006 & 2 \\
TESS $T$	&	5.345	&	0.007 & 3 \\
2MASS $J$	&	5.219	&	0.018	& 4\\
2MASS $H$	&	5.233	&	0.031	& 4 \\
2MASS {\it Ks}	&	5.206	&	0.018 & 4\\
WISE RSR $W1$	&	5.249	&	0.188 & 5\\
WISE RSR $W2$	&	5.092	&	0.071 &5 \\
\hline \hline
\end{tabular}
\tablefoot{{\bf References:} 1 -- \citet{Paunzen2015}, 2 -- \citet{GAIA2018}, 3 -- \citet{Paegert2021}, 4 -- \citet{Cutri2003}, 5 -- \citet{Cutri2021}}
\end{table}

\end{appendix}

\end{document}